\documentclass[sigconf]{acmart}
\usepackage{array}
\usepackage{multirow}

\newcommand{\pluto}{\textsc{Pluto}}
\newcommand{\gpluto}{g\textsc{Pluto}}
\newcommand{\OG}{\textsc{OpenGadget3}}
\newcommand{\gadget}{\textsc{Gadget-2}}
\newcommand{\ramses}{\textsc{Ramses}}
\newcommand{\ipic}{i\textsc{PIC3D}}
\newcommand{\changa}{\textsc{ChaNGa}}
\newcommand{\bhac}{\textsc{BHAC}}
\newcommand{\fil}{\textsc{FIL}}
\newcommand{\grace}{\textsc{GRACE}}

\AtBeginDocument{%
  }

\copyrightyear{2025}
\acmYear{2025}
\setcopyright{rightsretained}
\acmConference[CF Companion '25]{22nd ACM International Conference on Computing Frontiers}{May 28--30, 2025}{Cagliari, Italy}
\acmBooktitle{22nd ACM International Conference on Computing Frontiers (CF Companion '25), May 28--30, 2025, Cagliari, Italy}\acmDOI{10.1145/3706594.3728892} \acmISBN{979-8-4007-1393-4/2025/05}

\begin{document}

\title{EuroHPC SPACE CoE: Redesigning Scalable Parallel Astrophysical Codes for Exascale}
\subtitle{Invited Paper}


\author{Nitin Shukla, Alessandro Romeo, Caterina Caravita}
\affiliation{
  \institution{CINECA}
  \city{Bologna}
  \country{Italy}}
\email{n.shukla@cineca.it}

\author{Lubomir Riha, Ondrej Vysocky, Petr Strakos, Milan Jaros, João Barbosa, Radim Vavrik}
\affiliation{
  \institution{IT4Innovations, VSB-TU Ostrava}
  \city{Ostrava}
  \country{Czech Republic}}
\email{lubomir.riha@vsb.cz}

\author{Andrea Mignone, Marco Rossazza, Stefano Truzzi, Vittoria Berta, Iacopo Colonnelli, Doriana Medi\'c}
\affiliation{
  \institution{University of Turin} 
  \city{Turin}
  \country{Italy}}
\email{andrea.mignone@unito.it}

\author{Elisabetta Boella, Daniele Gregori}
\affiliation{
  \institution{E4 COMPUTER ENGINEERING SpA}
  \city{Scandiano}
  \country{Italy}}
\email{elisabetta.boella@e4company.com}

\author{Eva Sciacca, Luca Tornatore, Giuliano Taffoni}
\affiliation{
  \institution{INAF} 
  \city{Catania, Trieste}
  \country{Italy}}
\email{eva.sciacca@inaf.it}

\author{Pranab J Deka, Fabio Bacchini, Rostislav-Paul Wilhelm}
\affiliation{
  \institution{KU Leuven}
  \city{Leuven}
  \country{Belgium}}
\email{pranab.deka@kuleuven.be}

\author{Georgios Doulis, Khalil Pierre, Luciano Rezzolla}
\affiliation{
  \institution{Goethe-Universität}
  \city{Frankfurt}
  \country{Germany}}
\email{gdoulis@itp.uni-frankfurt.de}

\author{Tine Colman, Beno\^it Commer\c con}
\affiliation{
  \institution{CRAL, CNRS, ENS Lyon}
  \city{Lyon}
  \country{France}}
\email{tine.colman@cnrs.fr}

\author{Othman Bouizi, Matthieu Kuhn, Erwan Raffin, Marc Sergent}
\affiliation{
  \institution{CEPP, Eviden}
  \country{France}}
\email{erwan.raffin@eviden.com}

\author{Robert Wissing}
\affiliation{
  \institution{University of Oslo}
  \city{Oslo}
  \country{Norway}}
\email{robertwi@astro.uio.no}

\author{Guillermo Marin}
\affiliation{
  \institution{Barcelona Supercomputing Center} 
  \city{Barcelona}
  \country{Spain}}
\email{guillermo.marin@bsc.es}

\author{Gino Perna, Marisa Zanotti}
\affiliation{
  \institution{ENGINSOFT SpA} 
  \city{Trento}
  \country{Italy}}
\email{g.perna@enginsoft.com}

\author{Klaus Dolag, Geray S. Karademir}
\affiliation{
  \institution{Ludwig-Maximilians-Universität}
  \city{Munich}
  \country{Germany}}
\email{karademir@usm.lmu.de}

\author{Sebastian Trujillo-Gomez}
\affiliation{
  \institution{Heidelberg Institute for Theoretical Studies} 
  \city{Heidelberg}
  \country{Germany}}
\email{sebastian.trujillogomez@h-its.org}

\renewcommand{\shortauthors}{Shukla et al.}

\begin{abstract}
High Performance Computing (HPC) based simulations are crucial in Astrophysics \& Cosmology (A\&C), helping scientists investigate and understand complex astrophysical phenomena. Taking advantage of exascale computing capabilities is essential for these efforts. However, the unprecedented architectural complexity of exascale systems impacts legacy codes. The SPACE Centre of Excellence (CoE) aims to re-engineer key astrophysical codes to tackle new computational challenges by adopting innovative programming paradigms and software (SW) solutions. SPACE brings together scientists, code developers, HPC experts, hardware (HW) manufacturers, and SW developers. This collaboration enhances exascale A\&C applications, promoting the use of exascale and post-exascale computing capabilities. Additionally, SPACE addresses high-performance data analysis for the massive data outputs from exascale simulations and modern observations, using machine learning (ML) and visualisation tools. The project facilitates application deployment across platforms by focusing on code repositories and data sharing, integrating European astrophysical communities around exascale computing with standardised SW and data protocols. 
\end{abstract}






\maketitle

\section{The SPACE CoE and its flagship codes}
HPC-based numerical simulations are essential for modelling the physical processes shaping our universe and thus advancing A\&C knowledge. HPC is also critical for processing modern simulation outputs, requiring exascale computing to enable high-resolution, reliable analyses.
The transition to exascale computing presents challenges due to the complexity of modern HPC architectures, which require significant updates to existing simulation and data analysis codes. In response, European CoEs were established under Horizon Europe to enhance and scale parallel codes for exascale performance, fostering collaboration between academia, industry, and technology providers.
One such initiative is the Scalable Parallel Astrophysical Codes for Exascale (SPACE) CoE\footnote{SPACE: \url{https://www.space-coe.eu/}}, which aims to adapt and optimise European A\&C simulation codes for exascale systems. Supported by various European nations and organisations, SPACE promotes co-design activities, knowledge sharing, and the development of advanced computational techniques to maintain Europe’s leadership in scientific research.
SPACE focuses on improving scalability, energy efficiency, data processing, and visualisation capabilities, as well as the collaboration with ML and HPC experts to ensure that upgraded codes meet current and future computational demands in A\&C research.




\subsection{\pluto}
The \pluto~code delivers a modular, multiphysics and multi-al\-go\-rithm framework for simulating astrophysical flows in the presence of high-Mach number flows.
It allows independent selection of various hydrodynamic modules and algorithms to accurately model Newtonian hydrodynamics (HD), relativistic HD (RHD), magnetohydrodynamic (MHD), relativistic MHD (RMHD), and resistive relativistic MHD fluids (ResRMHD). 
The modular design is based on a robust framework for integrating hyperbolic conservation laws by means of modern Godunov-type shock-capturing schemes to ensure high accuracy. 
Generally speaking, these methods comprise three steps: a
reconstruction routine followed by the solution of a Riemann problem at zone edges, and a final update stage.

\begin{figure}[t]
    \centering
    \includegraphics[width = 0.24\textwidth]{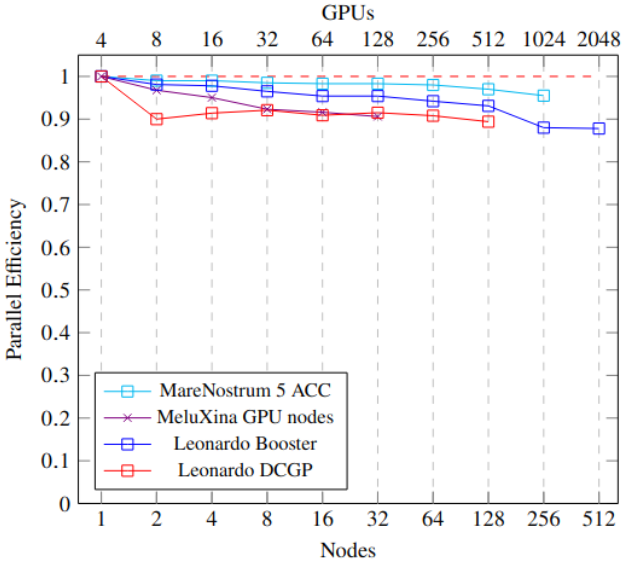}
    \includegraphics[width = 0.23\textwidth]{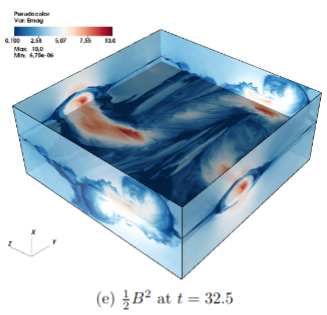}
    \caption{
    \gpluto~weak scaling results for the 3D Orszag-Tang on MareNostrum 5 ACC, MeluXina GPU nodes, and Leonardo (Booster and DCGP) (left). Snapshots of the magnetic energy density ($\frac{1}{2}B^2$) at time t = 32.5 (right). Here,
    time is expressed in units of the light-crossing time of the sheet length. 
    }
    \label{fig:pluto}
\end{figure}

The new upcoming version of the \pluto~ code, \gpluto\footnote{\gpluto: \url{https://gitlab.com/PLUTO-code/gPLUTO}}, has been specifically redesigned to support computations on Graphics Processing Units (GPUs) at the forefront of the exascale era.
More specifically, \gpluto~has been entirely rewritten in \verb|C++|, and relies on the \verb|OpenACC| programming model to provide acceleration up to thousands of GPUs, showing excellent parallel performance.
\gpluto~maintains as much backward compatibility as possible when compared to its predecessor and a very similar user interface. \verb|MPI| communications use non-blocking \verb|MPI| calls and asynchronous data exchange. 
The development of \gpluto~involved numerous modifications to optimise parallelisation and memory access, along with a transition from \verb|C| to \verb|C++|. This transition introduced classes for multidimensional arrays and function templates. Implementation details are described in \citep{berta_2024a} and \citep{mignone2024}. The left panel in Fig.~\ref{fig:pluto} shows weak scaling tests of the code conducted on CPU and GPU partitions of three different HPC platforms: MareNostrum 5 Accelerated Partition (GPU), Leonardo Booster (GPU), MeluXina GPU nodes (GPU), and Leonardo DCGP (CPU). 
The benchmark executed is the 3D version of the well-known Orszag-Tang test. Recently, \gpluto~has been used to perform 4th-order accurate 3D numerical simulation of magnetic reconnection, triggered by the tearing instability within a resistive relativistic MHD (ResRMHD) framework  \citep{Berta2024b} (Fig.~\ref{fig:pluto} right panel). These computations took advantage of the Leonardo Booster, enabling high-resolution results that would be otherwise unattainable within reasonable timeframes.


\subsection{\OG}
\begin{figure}[t]
    \centering
    \includegraphics[width = 0.28\textwidth]{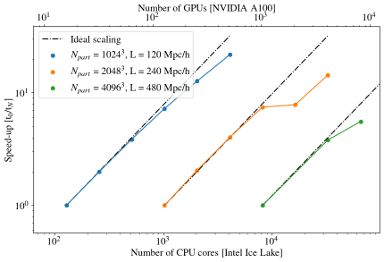}
    \includegraphics[width = 0.19\textwidth]{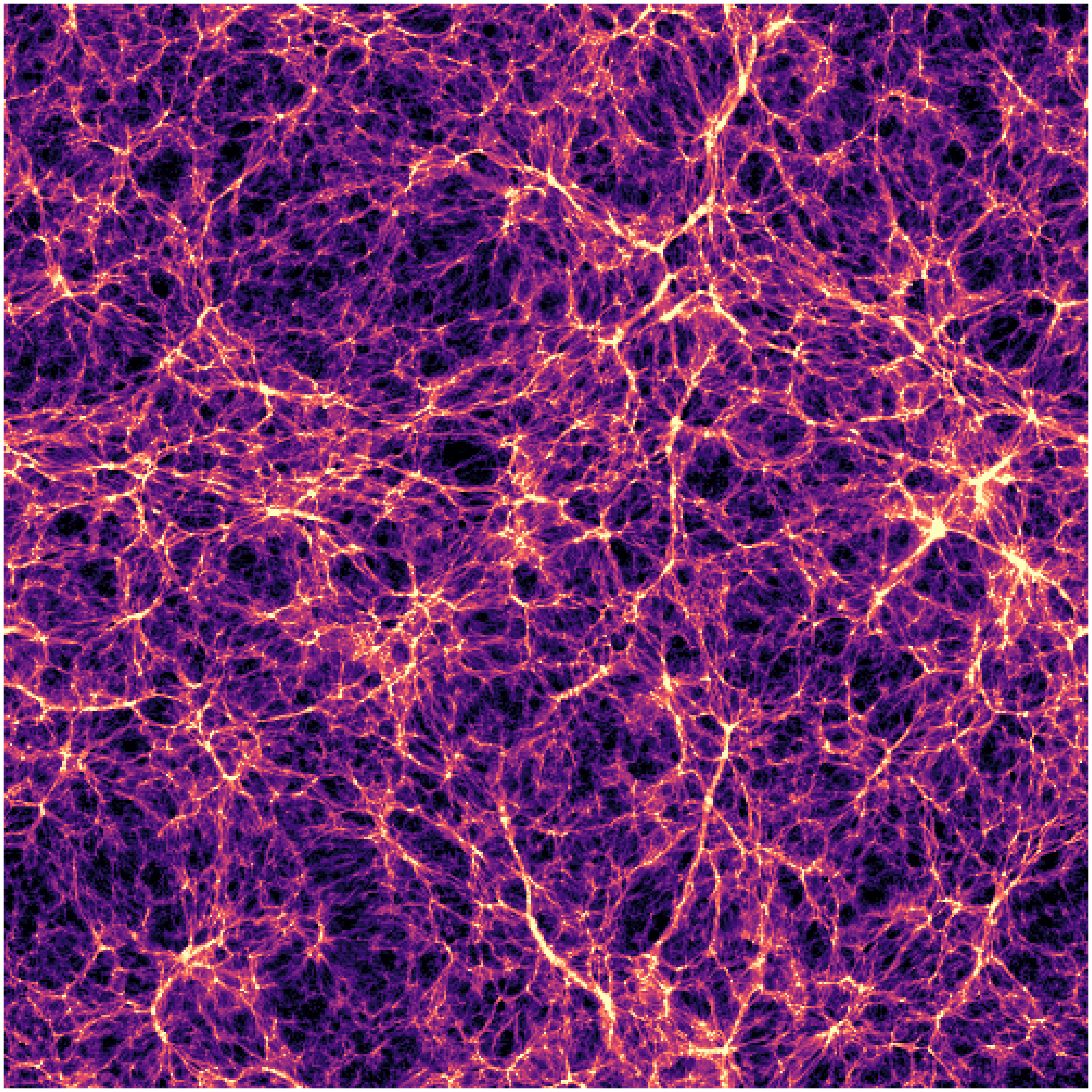}
    \caption{
    \OG~strong scaling results of different gravity-only boxes on the Leonardo Booster using three test cases (left). Visualisation of Box3 from the Magneticum Pathfinder simulation set (right). The shown region spans a total size of $\sim180$ Mpc and contains $\sim3.5*10^{8}$ dark matter, gas, star, and black hole particles (image credit: B. Seidel).
    }
    \label{fig:og}
\end{figure}

\OG\footnote{\OG: \url{https://gitlab.lrz.de/MAGNETICUM/Hydro-OpenGadget3}} is a collisionless N-Body/Lagrangian cosmological code that uses the smoothed particle HD (SPH) computational method to describe the motion of fluids in addition to gravitational forces, which are calculated using a tree structure. The code allows simulations in a full cosmological context, i.e. accounting for an expanding background and the presence of matter, both ``dark” and baryonic (ordinary matter), and dark energy. In addition to the gravitational problem, it also simulates the evolution of baryonic matter, accounting for HD and various physical effects such as radiative cooling, star formation, energy feedback, radiative transfer, magnetic fields, and more. Although the full cosmological context is often the default choice (see right panel of Fig.~\ref{fig:og} for an example), having a non-expanding background and a setup with only dark or baryonic matter is equally possible.

\OG~evolved from the publicly available \gadget~code \cite{Springel2005}. The code is written in \verb|C/C++| and uses a hybrid parallelisation (\verb|MPI+OpenMP|). \OG~has been significantly improved compared to its base version, for example, by adding a new state-of-the-art SPH implementation \cite{Beck2016}, a meshless finite mass solver \cite{Groth23}, and \verb|OpenACC| support for running on multiple GPUs \cite{Ragagnin2020}. As shown in the left panel of Fig.~\ref{fig:og}, the code using the \verb|OpenACC| implementation is scaling well up to a few thousand GPUs. Based on this, we are also implementing \verb|OpenMP| offloading, and have experimented with a new strategy of finding neighbours and walking the gravity tree. By coalescing the walk for bunches of particles that belong to the same space region, we consequently synchronise threads and avoid memory divergence. Calculating the gravitational force via direct summation of all the particles within such a bunch of particles avoids branching and conditionals, thus enhancing vectorisation. A proof of concept shows an additional $\sim 10 \times$ speedup compared to the currently employed approach.


\subsection{\ramses}
\begin{figure}[t]
    \centering
    \includegraphics[width=0.53\columnwidth]{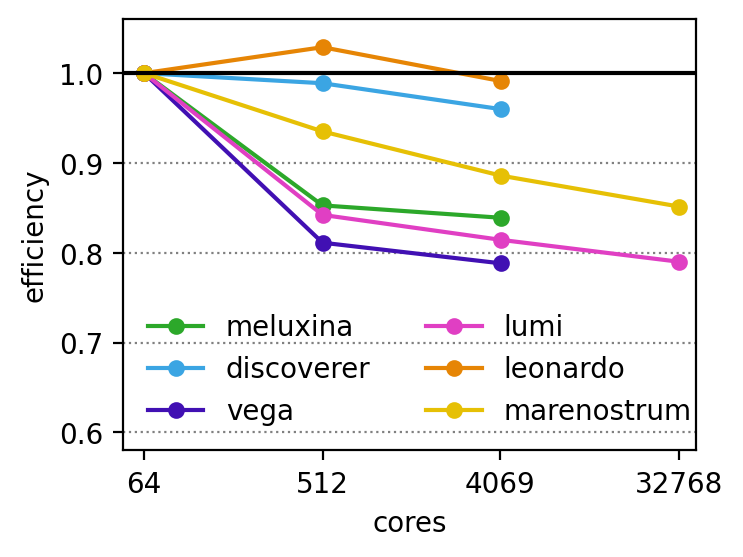}
    \includegraphics[width=0.40\columnwidth]{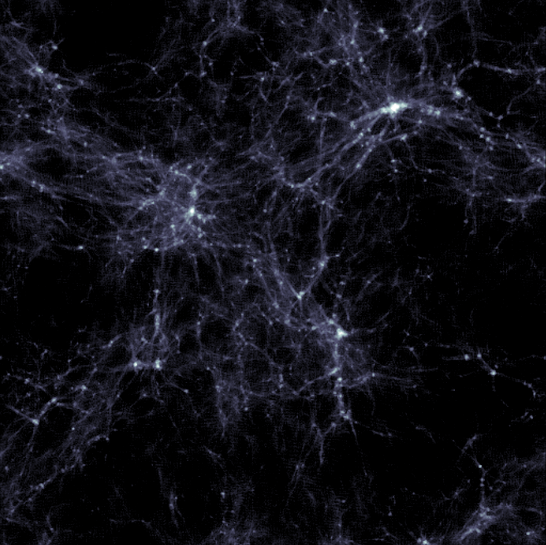}
    \caption{
    Weak scaling of \ramses~(before optimisation) on the CPU partition of several EuroHPC systems for the Sedov test case with 128$^3$ cells per core (left). Projected image of the cosmological volume test case (right).}
    \label{fig:ramses}
\end{figure}

\ramses\footnote{\ramses: \url{https://git-cral.univ-lyon1.fr/hpc/space/ramses}} is an adaptive mesh refinement (AMR) code used to study astrophysical fluid dynamics and the formation of structures in the Universe \citep{Teyssier2002}.
It is based on an oct-tree structure, where parent cells are refined into children cells on a cell-by-cell basis following some user-defined criteria.
It couples an Eulerian solver for gas dynamics (hydro) with a particle-mesh (PM) method for collisionless components such as dark matter. 
The diversity of physics modules included in the code, such as MHD, radiative transfer, self-gravity, and various star formation models, allow for the detailed simulation of complex astrophysical systems ranging from stellar interiors to large cosmological volumes.
We focus our efforts on improving performance on the main modules (hydro, gravity, PM, and AMR), as probed by three selected test cases: a Sedov blast wave, a cosmological box (Fig.~\ref{fig:ramses}, right panel), and an isolated galaxy.
The left panel of Fig.~\ref{fig:ramses} illustrates the weak scaling of the Sedov test.

\ramses~is written in \verb|Fortran90| and parallelised with \verb|MPI| using domain decomposition.
Internal vectorisation is already excellent. 
To further improve scaling, we focus on tackling issues related to \verb|MPI| communications.
A key limitation when scaling \ramses~to a larger number of processes is the increasing surface-to-volume ratio of the \verb|MPI| domains. This results in a significant memory overhead due to boundary duplication between neighbouring domains. To address this, we are integrating \verb|OpenMP| to implement a hybrid parallelisation strategy, reducing the number of \verb|MPI| domains while increasing their volume. This approach improves memory efficiency by a factor of 10 and decreases \verb|MPI| communication overhead, resulting in better execution times and enhanced scalability.



\subsection{\ipic}

\begin{figure}[t]
    \centering
    \includegraphics[width = 0.16\textwidth]{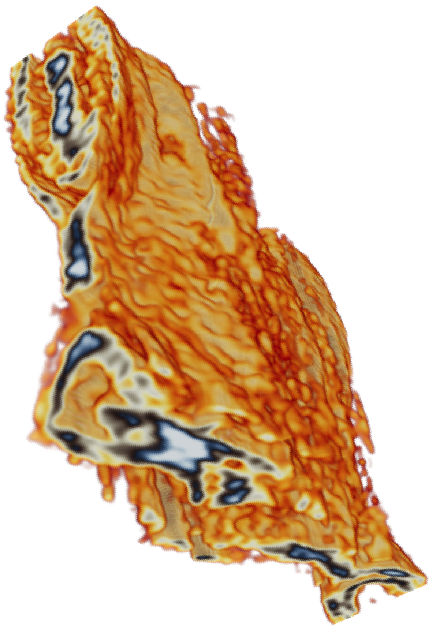}
    \includegraphics[width = 0.31\textwidth]{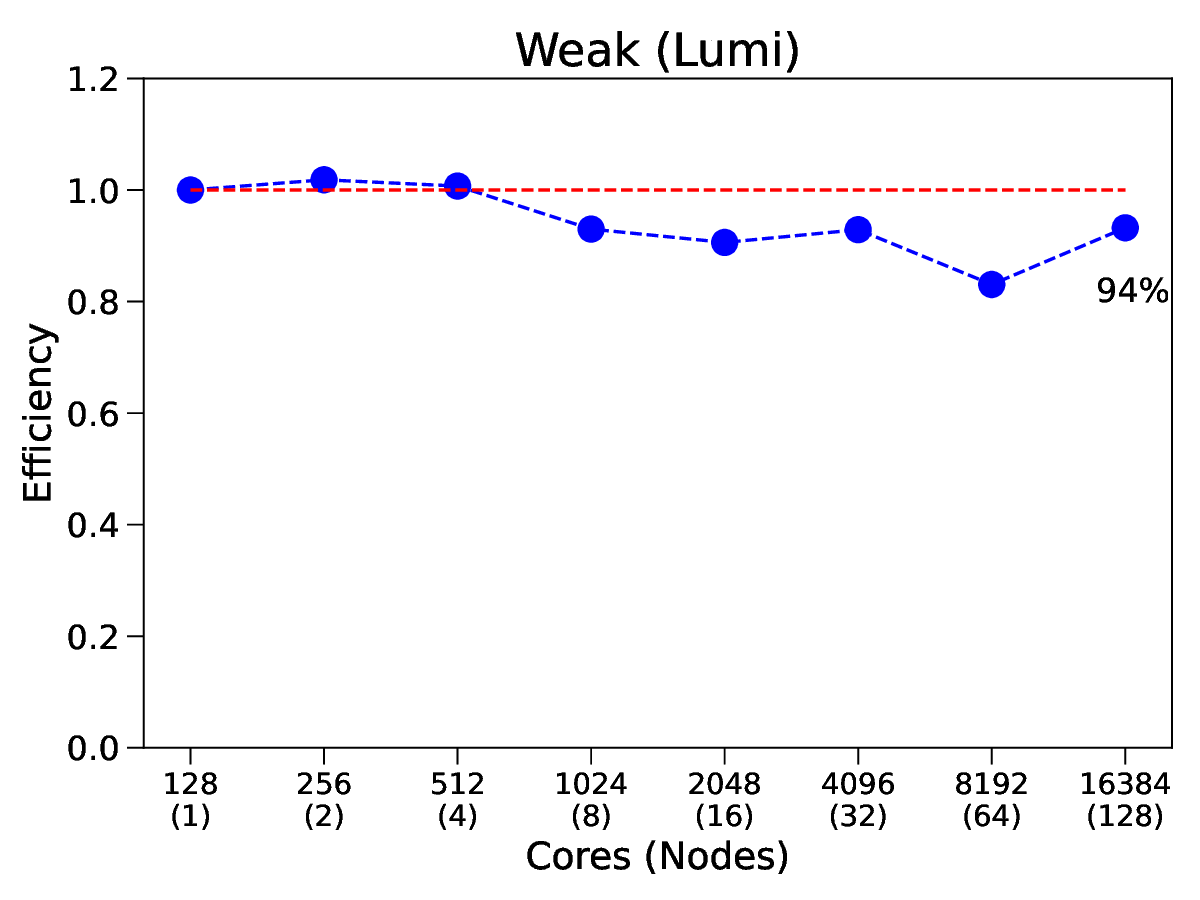}
    \caption{Current-density distribution of a high-resolution 3D simulation of relativistic magnetic reconnection with \ipic~(left). Weak scaling of \ipic~on LUMI-C (right).}
    \label{fig:ipic3d}
\end{figure}

The Implicit Particle-in-Cell 3D (\ipic\footnote{\ipic: \url{https://github.com/Pranab-JD/iPIC3D-CPU-SPACE-CoE}}) code \citep{Markidis2010} is a Particle-in-Cell (PIC) simulation tool developed primarily to study plasma dynamics on kinetic scales (see the left panel of Fig.~\ref{fig:ipic3d} for an example of simulation). The individual (macro)particles used to represent plasma particles are evolved in a Lagrangian framework whereas the moments (plasma density, current, etc) and the self-consistent electric and magnetic fields are tracked on an Eulerian grid. The three main kernels of \ipic~are (a) Particle Mover, (b) Moment Gatherer, and (c) Field Solver. Due to the implicit nature of the underlying algorithm, as opposed to explicit PIC methods, unresolved scales do not lead to numerical instabilities. This enables the use of time steps and spatial grid sizes that are 10 to 100 times larger than those typically required in traditional explicit PIC codes.

In collaboration with KTH Sweden, we have developed a new version of the code that offers GPU support with \verb|CUDA| (for NVIDIA GPUs) and \verb|HIP| (for AMD GPUs), nonblocking asynchronous communication across \verb|MPI| subdomains, and \textit{in-situ} visualisation using Paraview/Catalyst. The right panel of Fig.~\ref{fig:ipic3d} illustrates the weak scaling of the code on LUMI-C, demonstrating a parallel efficiency of over 90\%.

We have recently implemented the Energy Conserving Semi-Implicit Method (ECSIM)  \citep{Lapenta17,Lapenta_Gonzalez-Herrero_Boella_2017} in this new version of the code which enables us to conserve energy of the system exactly up to machine precision. We have also implemented the relativistic semi-implicit method (RelSIM) \citep{Bacchini23}, which is the relativistic counterpart of ECSIM. At present, the moment gatherer module is the most expensive part of the computation, encompassing $\sim 80\%$ of the overall runtime. We aim to further optimise this module via vectorisation.
After optimisation, we will implement ECSIM and RelSIM algorithms in the GPU code. This will follow ongoing efforts to scale these algorithms, along with \ipic, to several thousand GPUs, potentially paving the way for exascale computing.


\subsection{\changa}

\begin{figure}[t]
    \centering
    \includegraphics[width = 0.20\textwidth]{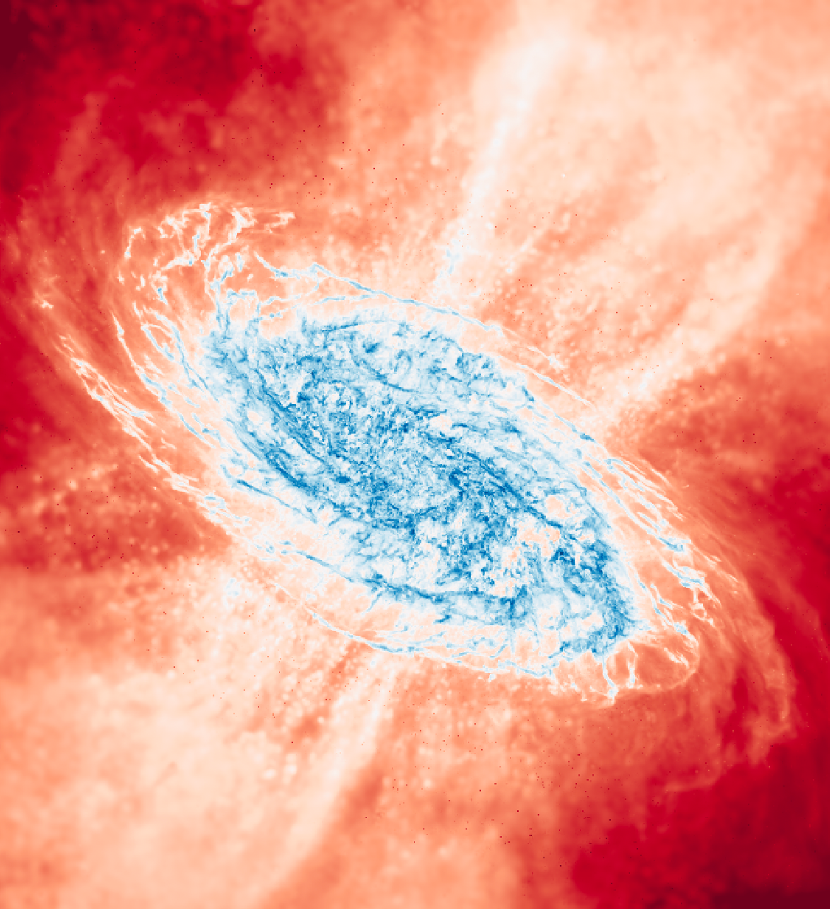}
    \includegraphics[width = 0.27\textwidth]{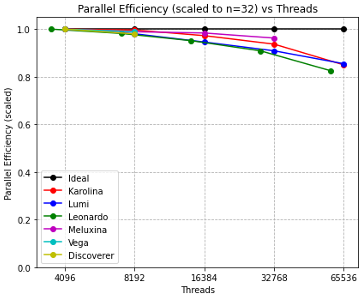}
    \caption{Magnetic field strength distribution from a high-resolution galaxy simulation performed with \changa~(left). Weak scaling performance of \changa~across multiple EuroHPC supercomputers (right).}
    \label{fig:changa}
\end{figure}

\changa\footnote{\changa: \url{https://github.com/N-BodyShop/changa}} is an N-body and smoothed particle MHD (SPMHD) code designed to simulate a wide range of astrophysical systems \cite{ChaNGa1, ChaNGa2} (see Fig.~\ref{fig:changa} left panel for an example). Building upon the gravity and SPMHD algorithms of gasoline~\cite{ChaNGaSPH} and pkdgrav~\cite{ChaNGaGrav}, \changa~leverages the \verb|Charm++| framework to deliver superior parallel scalability compared to its predecessors. 
This scalability stems from three key features of \verb|Charm++|. First, overdecomposition, which divides the computational workload into many more units (called chares or tree pieces) than available processors. Second, dynamic load balancing, where the \verb|Charm++| runtime system dynamically manages chares through continuous load-balancing strategies, evaluating and migrating tree pieces between processors during execution to maintain optimal load distribution. Third, asynchronous task-based execution model, in which computation is message-driven, with tasks triggered by asynchronous messages. This model enables overlapping computation and communication, reducing idle time and enhancing efficiency. The right panel of Fig.~\ref{fig:changa} demonstrates \changa~performance, showing excellent scaling across up to 65,536 cores on the CPU partitions of multiple EuroHPC supercomputers. \verb|Charm++| also provides support to execute \verb|CUDA| kernels on the GPU asynchronously and to manage data transfers between the CPU and GPU. \changa~has previously only ported its gravity module to GPU, which has shown excellent scaling up to $\approx 3,000$ nodes (1 GPU per node) on the Piz Daint supercomputer. Recently, we have begun enhancing \changa’s GPU module by offloading computationally intensive tasks to the GPU. As a first step, we have implemented a preliminary GPU offload for the radiative cooling module, achieving a speedups varying from $4 \times$ to $20 \times$ depending on the test case. Additionally, we have improved communication balance within \changa~by implementing tree-piece replication, which distributes tree pieces across multiple processors, preventing any single processor from being overwhelmed with messages.

\subsection{\bhac}
\begin{figure}[t]
    \centering
    \hspace{-4mm}\includegraphics[width = 0.27\textwidth]{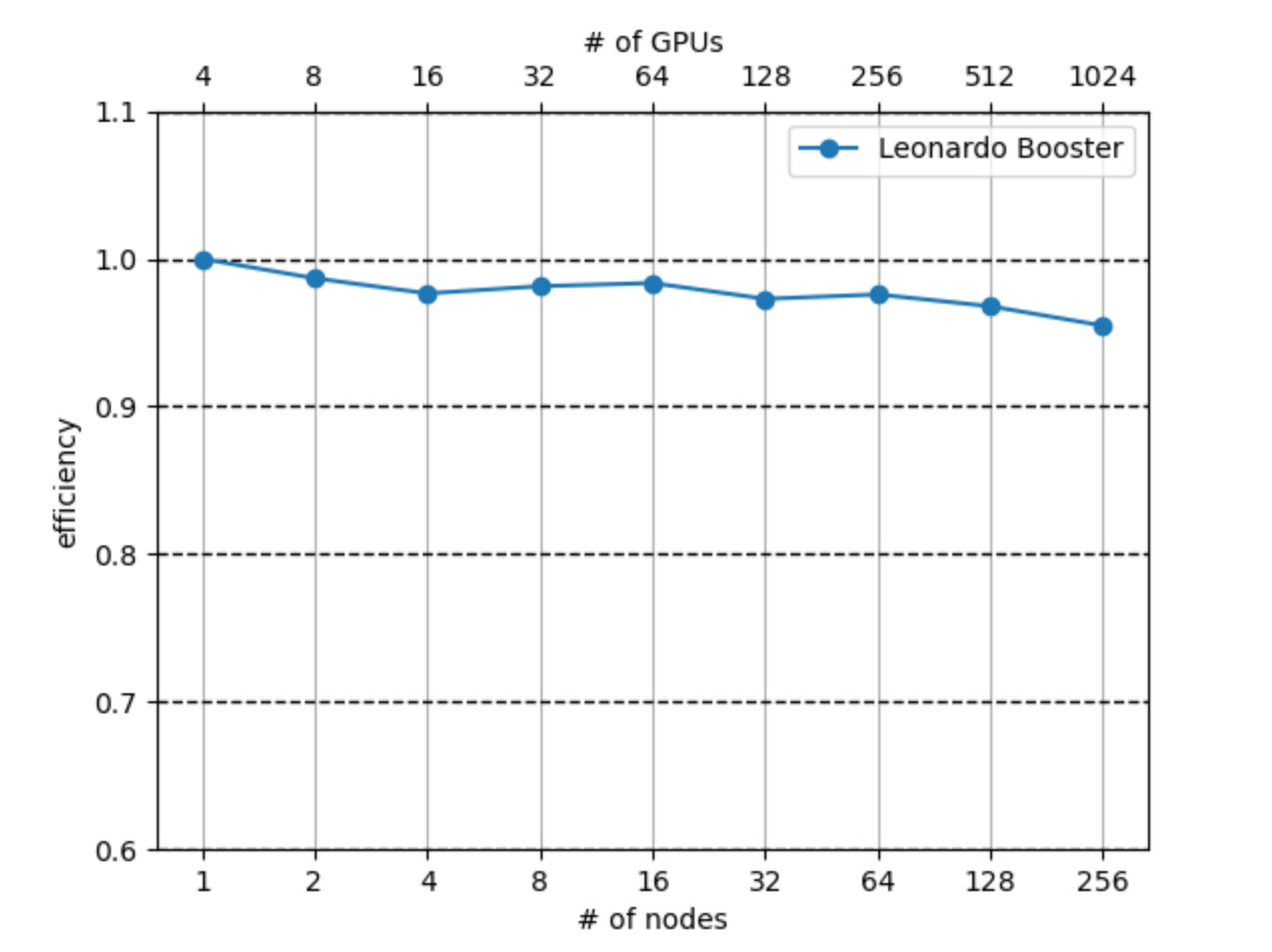}\hspace{-4mm}
    \includegraphics[width = 0.23\textwidth]{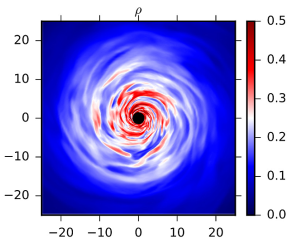}
    \caption{\bhac~weak scaling (left) on Leonardo Booster (GPU partition) for the 3D simulation of a magnetised torus around a Kerr black hole--z=0 plane shown (right).}
    \label{fig:bhac}
\end{figure}

The Black Hole Accretion Code (\bhac\footnote{\bhac: \url{https://gitlab.itp.uni-frankfurt.de/BHAC-release/bhac}}) is a multidimensional General RMHD (GRMHD) code that is mainly used to study accretion flows onto compact objects \cite{Porth_2017,Olivares_2019}. \bhac~has been designed to solve the GRMHD equations in arbitrary stationary space-times (Cowling approximation) exploiting AMR techniques with an oct-tree block-based approach provided by the \textsc{MPI-AMRVAC} framework\footnote{\textsc{MPI-AMRVAC}: \url{https://amrvac.org}}.
The code is second-order accurate and uses finite volume and high-resolution shock-capturing (HRSC) methods. Originally designed to study black hole (BH) accretion in ideal GRMHD, \bhac~has been extended to incorporate nuclear equations of state, neutrino leakage, charged and purely geodetic test particles, and non-black hole fully numerical metrics. In addition, a non-ideal resistive GRMHD module has been developed and implemented. 
\bhac’s results, after a general-relativistic ray-tracing (GRRT) post-processing, can be used to compute synthetic observable images of BH shadows and the surrounding accretion flows. These calculations are performed with the GRRT Black Hole Observations in Stationary Spacetimes (BHOSS) code \cite{Younsi_2020}. The GRMHD simulation data produced by \bhac~are used as input for BHOSS to produce accretion flow and BH shadow images. Currently, two modules have been ported to GPU using \verb|OpenACC| and are fully operational on multiple GPUs: the primitive reconstruction scheme and the Riemann solver. The former is the most resource demanding module of \bhac. The multi-GPU performance of \bhac~is presented in Fig.~\ref{fig:bhac} (left panel) where an approximate 95\% weak scaling efficiency up to 1,024 GPUs is achieved. As test case, a 3D simulation of a magnetised torus around a Kerr black hole--z=0 plane was used (right panel of Fig.~\ref{fig:bhac}). The current GPU-port yields a $\sim 20 \times$ speedup over the CPU version. Ongoing optimisations aim to boost performance further by improving parallel efficiency and reducing CPU-GPU data transfers.

\subsection{\fil/\grace}

\begin{figure}[t]
    \centering
    \includegraphics[width = 0.27\textwidth]{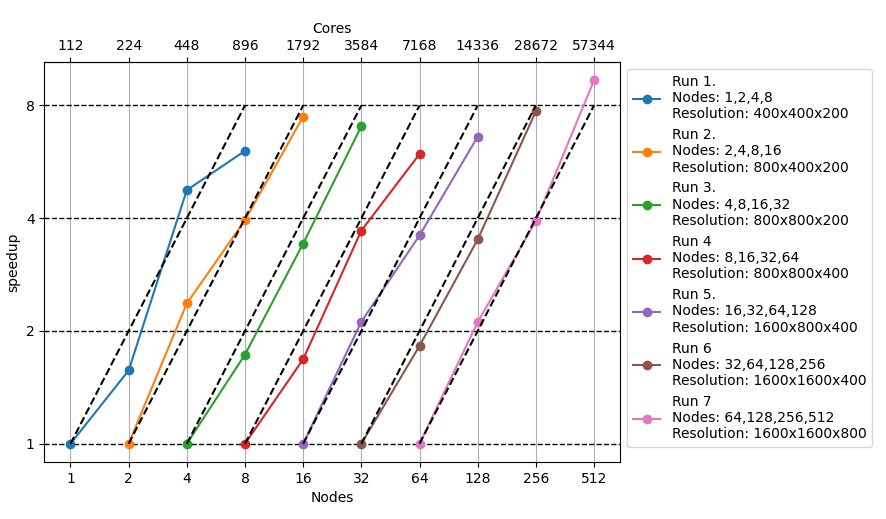}
    \includegraphics[width = 0.2\textwidth]{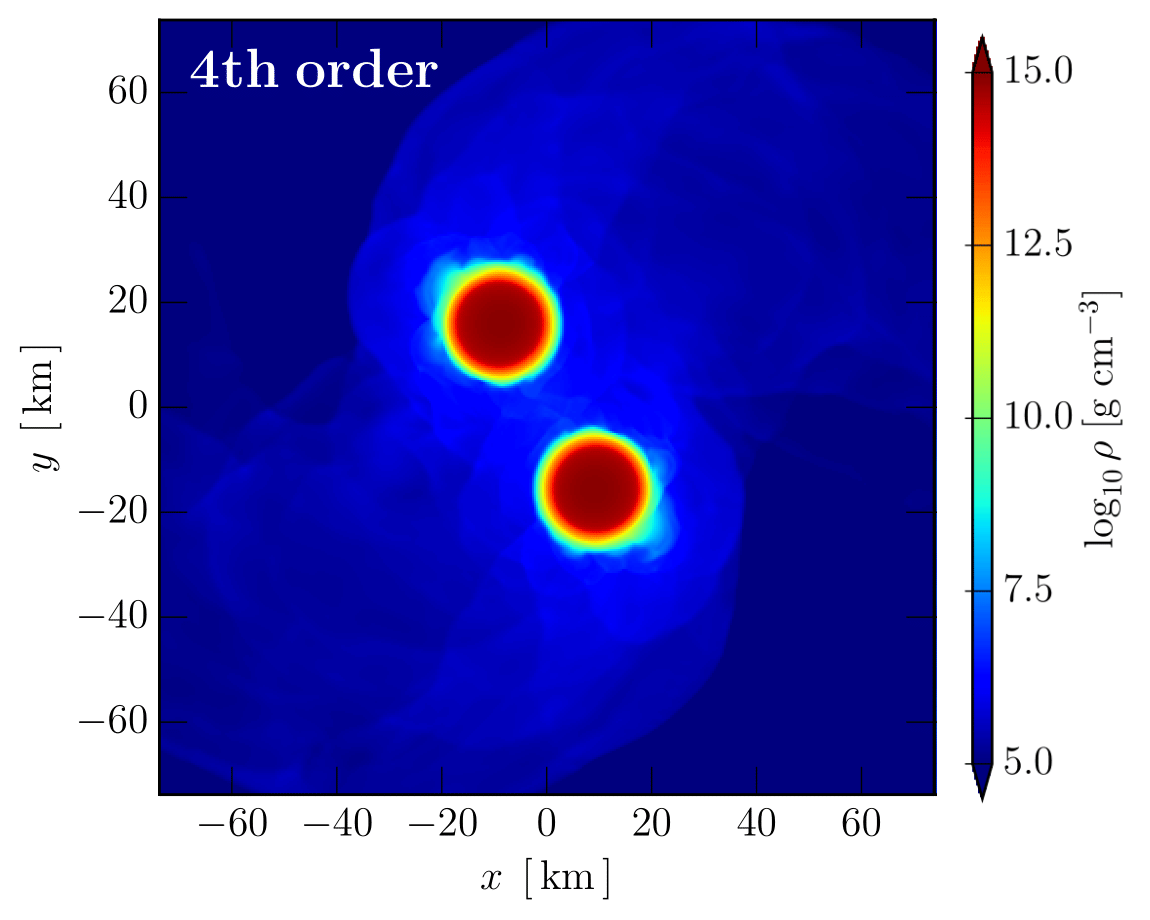}
    \caption{Strong and weak scaling of \fil~on Marenostrum (left). Density plot of BNS collision (right).}
    \label{fig:FIL1}
\end{figure}

\fil~is a GRMHD code capable of simulating relativistic fluids on a curved, dynamically evolving spacetime. This feature, which distinguishes \fil~from \bhac, makes it well-suited for modelling BH neutron star (BHNS) and binary neutron star (BNS) collisions (see the right panel of Fig.~\ref{fig:FIL1} for an example).
\fil~leverages the computational infrastructure of the Einstein Toolkit (ET), a codebase designed to support numerical relativity simulations. This includes features such as box-in-box AMR and fourth-order Runge-Kutta integration methods, both of which are employed by \fil.
\fil~is the successor to Illinois GRMHD, the first open-source GRMHD code for BNS simulations. With fourth-order finite difference methods and a tabulated Equation of State (EOS) interface, \fil~offers improved accuracy and enables exploration of the nuclear EOS parameter space.
\fil~is written in \verb|C++| and uses \verb|MPI| and \verb|OpenMP| for parallelisation. Strong and weak scaling tests of the code are shown in the left panel of Fig.~\ref{fig:FIL1}.

Work is currently underway to replace the functionalities of the ET framework used in \fil~with a new computational infrastructure called \grace. \grace~is a GRMHD codebase built on \verb|Kokkos| to leverage heterogeneous HW. This will allow \fil~to run on a multitude of GPU models. Unlike the current \fil~implementation, which is strictly based on Cartesian coordinates, \grace~supports multiple coordinate systems. This flexibility enables simulations to exploit problem-specific symmetries, improving computational efficiency.
\section{Profiling, Co-design \& Energy Efficiency}
\noindent\textbf{Profiling}. 
Optimisation of all SPACE codes is a key objective of the project. To enable focused optimisation and GPU acceleration of the most critical parts of the codes, a rigorous profiling campaign was done using the POP3 CoE\footnote{POP: Performance Optimisation and Productivity CoE: \url{https://pop-coe.eu/}} methodology and tools.

\noindent\textbf{Co-design}.
Another key objective of SPACE is to prepare code for emerging European technologies, as those developed within the EPI-SGA2\footnote{EPI: European Processor Initiative: \url{https://www.european-processor-initiative.eu/}} and EUPEX\footnote{EUPEX: European Pilot for Exascale: \url{https://eupex.eu/}} projects, by targeting the Rhea processor from SiPearl with High Bandwidth Memory (HBM) and ARM Neoverse~V1 cores. This involves close collaboration between application developers and HW/SW technology providers to facilitate knowledge exchange and mutual influence. Our investigation focuses on two primary performance tracks: (i) evaluating the impact of HBM versus Double Data Rate (DDR) memory, using Intel Sapphire Rapids HBM processors as a reference; and (ii) assessing the maturity of the ARM ecosystem, including compilers and third-party libraries, with a particular focus on the capability to compile code that efficiently leverages Scalable Vector Extension (SVE) for Single Instruction Multiple Data (SIMD) operations. In the absence of the Rhea processor, we currently use ARM-based platforms such as the NVIDIA Grace CPU for early development and testing. For example, Table~\ref{tab:performance_vecto} showcases the performance gains achieved by optimising the vectorisation of \ramses~using different compilers on both x86\_64 and AARCH64 architectures. While performance improvements were observed with all compilers, the optimal set of optimisations is dependent on both the compiler and the architecture. This dependency complicates the task of achieving portable code vectorisation.



\begin{table}[]
\footnotesize
\centering
\begin{tabular}{ |c|c|c|c|c|c|c|c| } 
 \hline
  & \multicolumn{3}{c|}{X86\_64} & \multicolumn{4}{c|}{AARCH64} \\ 
 \hline
 \textbf{Compiler} & GNU & IFORT & IFX & GNU  & GNU & ACFL & ACFL \\ 
  &  &  &  & NEON & SVE2 & NEON & SVE2 \\ 
 \hline
 \textbf{Time gain (\%)} & 3.7 & 1 & 9 & - & 9.6 & 2.7 & 4.1 \\ 
 \hline
\end{tabular}
\caption{Gains achieved on \ramses~hydro module by optimising vectorisation with different compilers and architectures.} 
\label{tab:performance_vecto} 
\end{table}

\noindent\textbf{Improving Energy Efficiency}. 
As energy awareness is becoming an essential focus for all data centre operators, all SPACE codes have been analysed in terms of energy efficiency on several HPC machines and modern HW platforms. We have investigated how to tune the performance knobs provided by different platforms (CPU core frequency (CF), CPU uncore frequency (UCF), and GPU streaming multiprocessor frequency) to minimise energy consumption when running SPACE codes. Table~\ref{tab:methodology:summary} shows that ARM-based NVIDIA Grace CPU delivers much better energy efficiency. In addition, by proper HW setup we can save up to 22\% of energy without runtime impact. For A100 GPUs, energy consumption can be reduced by up to 9\% with minimal runtime impact. For more details see public deliverables D2.2 and D2.4\footnote{SPACE deliverables: \url{https://www.space-coe.eu/deliverables.php}}.

\begin{table}[t]
    \footnotesize
    \centering
    \begin{tabular}{|>{\raggedright\arraybackslash}p{0.12\linewidth}  |>{\centering\arraybackslash}p{0.22\linewidth}|>{\centering\arraybackslash}p{0.21\linewidth}  |>{\centering\arraybackslash}p{0.15\linewidth}  |>{\centering\arraybackslash}p{0.10\linewidth}|}
        \hline
         HW & CPU or GPU frequency
[GHz]& Energy efficiency
[MFLOPs/W]& Node energy savings& Runtime impact\\
        \hline
         \multirow{3}{\linewidth}{Nvidia A100 GPU}& default & - & --& 100\,\%\\
         \cline{2-5}
         & 1.29& - & -6\,\%& 103\,\%\\
         \cline{2-5}
         & 1.11& - & -9\,\%& 113\,\%\\
        \hline
         \multirow{3}{\linewidth}{Intel  Xeon 9468 w. HBM}& default & 376& --& 100\,\%\\
         \cline{2-5}
         & CF 2.6; UCF 1.4& 398& -4\,\%& 101\,\%\\
         \cline{2-5}
         & CF 2.6; UCF 1.0& 411& -6\,\%& 105\,\%\\
        \hline
         \multirow{3}{\linewidth}{Nvidia Grace CPU} & default & 628& --& 100\,\%\\
         \cline{2-5}
         & CF 2.9\,GHz & 805& -22\,\%& 101\,\%\\
         \cline{2-5}
         & CF 2.1\,GHz & 964& -35\,\%& 122\,\%\\
        \hline
    \end{tabular}
    \caption{Energy efficiency analysis results for \pluto~code.}
    \label{tab:methodology:summary}
\end{table}







\section{Extreme Data Processing and Analysis}

One SPACE CoE objective is to integrate data analysis techniques with exascale A\&C applications to enhance scientific discoveries from numerical simulations. \emph{Extreme data processing and analysis} addresses this topic by developing prototype frameworks based on workflow engines to design modular HPC applications \cite{COLONNELLI2024} and on visualisation and ML tools detailed hereafter. 

\noindent\textbf{High Performance Visualisation}. 
State-of-the-art large-scale A\&C simulations generate petabytes of data, capturing the evolution of millions of objects across 3D space and numerous physical parameters. Developing visualisation systems for pre-exascale architectures is challenging due to HW/SW compatibility, data management, scalability, energy efficiency, and resource constraints, making development and porting processes complex and requiring careful, system-specific optimisation.
SPACE is developing three visualisation strategies to address the aforementioned challenges, which are relevant to the A\&C community \cite{tuccari2025} and potentially applicable to simulation results from other scientific domains: 
1) A novel solution for \textit{in-situ} visualisation using Hecuba\footnote{Hecuba: \url{https://github.com/bsc-dd/hecuba}}, where the analysis and visualisation run concurrently with the simulation, bypassing the need to store the full outputs and permitting the visualisation during runtime.
2) A workflow to produce cinematic visualisations \cite{cfd_vis_workflow} of large volumetric datasets in Blender\footnote{Blender: \url{https://www.blender.org/download/}} with high-level of control over the image quality, applied to SPMHD data from \changa. 
3) The adaptation of VisIVO\footnote{VisIVO: \url{https://visivo.readthedocs.io/}} modular applications \cite{sciacca2015integrated} for high-performance data visualisation to efficiently exploit (pre-)exascale HPC systems. For this approach, the workflow management system StreamFlow \cite{StreamFlow} is adopted in order to improve portability and reproducibility of the workflow systems by simplifying the definition, execution, and management of the different visualisation tasks.

\noindent\textbf{ML for Astrophysics}.
Simulations are sophisticated models used to interpret observations and make testable predictions as an integral part of the scientific method. 
However, their growing size and complexity is becoming a barrier to their own interpretability.
Exascale A\&C simulations will produce outputs that will be very challenging for humans to interpret and analyse. 
SPACE is leveraging powerful scalable ML methods to maximise the scientific discovery potential of these expensive simulations with a focus on exploration, interpretation, and inference.   
In addition to facilitating the interpretation and analysis of simulation outputs, we have identified the potential to integrate ML models directly into codes at runtime. This approach offers a promising path to efficiently and accurately incorporate physical processes that are currently too computationally expensive to model in large-scale simulations. SPACE is developing three ML applications to enhance the scientific workflows of exascale A\&C simulations. The first one consists of a modular framework for interactive explorative access and knowledge discovery in arbitrarily large cosmological simulations using Representation Learning \cite{spherinator}. The second application complements this with a general inference tool that allows cosmological simulations to predict the physical properties and evolution of observed cosmic structures based on observables. The last application aims to develop a surrogate ML model to include costly radiative transfer effects in the largest cosmological simulations by pretraining using detailed calculations.    

\section {Services For the Community}
Transition to exascale HPC demands an up-to-date approach to code writing and integration. 
It is within this context that the following services have been implemented: services for continuous integration, issue tracking, and data archiving form the foundation of the modern development and research infrastructure. Also, a new vision of continuous deployment is under evaluation in order to be compatible with national computing centres and their security policies and to deal with target machine recipes in a much easier way.
The SPACE CoE is defining I/O standards and data models for the A\&C community. The goal is to enable data exchange between compatible simulation codes and make outputs accessible to the broader community. A standardised metadata model is also being developed to describe simulation results, supporting discovery and reuse. This model follows the Findable, Accessible, Interoperable, Reusable (FAIR) principles and builds on the work of the International Virtual Observatory Alliance (IVOA).

\section{Training activities}

A key focus of the SPACE CoE is delivering high-quality training to support the broader scientific and industrial community. To this end, SPACE organises a diverse range of events, including webinars, workshops, tutorials, schools, and hackathons, targeted at early-career researchers and professionals. These activities cover topics such as A\&C codes, dataset discovery, data processing, ML, HPC visualisation, and energy efficiency in HPC. Training content also includes updates on exascale development from SPACE code developers and HPC experts. All materials are publicly available via the SPACE CoE website and YouTube channel. SPACE actively collaborates with initiatives like EuroCC and CASTIEL (Training Sprint), alongside other CoEs and host institutions, to expand its reach. Events are promoted through the website, LinkedIn, newsletters, and mailing lists to engage the HPC community and potential users. 

\section{Summary}
The SPACE CoE has recently passed its two-year milestone and is now fully engaged in application optimisation and porting efforts for both current and future EuroHPC systems. Our results demonstrate strong code scalability and active porting to advanced architectures, ensuring long-term sustainability and performance. Extensive work is also underway in energy efficiency, profiling, optimisation, and extreme-scale data processing, analysis, and visualisation. For the latest updates and detailed information, please visit our project website\footnote{SPACE: \url{https://www.space-coe.eu/}}. 

\begin{acks}
SPACE CoE is funded by the European Union. It has received funding from the European High Performance Computing Joint Undertaking and from Belgium, the Czech Republic, France, Germany, Greece, Italy, Norway, and Spain under grant agreement No. 101093441.

We acknowledge ISCRA for awarding this project access to the Leonardo supercomputer and the EuroHPC Joint Undertaking for granting us access to the Leonardo, MareNostrum and MeluXina supercomputer through EuroHPC 
Benchmark and Development Access calls.
\end{acks}

\bibliographystyle{ACM-Reference-Format}
\bibliography{sources}

@String{Computing = "Computing" }

@String{Computer = "{IEEE} Computer" }

@String{Springer = "Springer-Verlag" }

@article{Markidis2010,
      title = {{Multi-scale simulations of plasma with iPIC3D}},
    journal = {Math. Comput. Simul.},
     volume = {80},
      pages = {1509-1519},
       year = {2010},
        doi = {10.1016/j.matcom.2009.08.038},
     author = {S. Markidis and G. Lapenta and Rizwan-uddin}
}

@ARTICLE{Lapenta17,
       author = {{Lapenta}, G.},
        title = "{Exactly energy conserving semi-implicit particle in cell formulation}",
      journal = {J. Comput. Phys.},
         year = 2017,
       volume = {334},
        pages = {349-366},
          doi = {10.1016/j.jcp.2017.01.002},
       adsurl = {https://ui.adsabs.harvard.edu/abs/2017JCoPh.334..349L}
}

@article{Lapenta_Gonzalez-Herrero_Boella_2017, title={Multiple-scale kinetic simulations with the energy conserving semi-implicit particle in cell method}, volume={83}, DOI={10.1017/S0022377817000137}, number={2}, journal={Journal of Plasma Physics}, author={Lapenta, Giovanni and Gonzalez-Herrero, Diego and Boella, Elisabetta}, year={2017}, pages={705830205}}

@ARTICLE{Bacchini23,
       author = {{Bacchini}, F.},
        title = "{RelSIM: A Relativistic Semi-implicit Method for Particle-in-cell Simulations}",
      journal = {ApJS},
         year = 2023,
       volume = {268},
       number = {2},
          doi = {10.3847/1538-4365/acefba},
       adsurl = {https://ui.adsabs.harvard.edu/abs/2023ApJS..268...60B}
}

@ARTICLE{tuccari2025,
       author = {Tuccari, N. and 
                 Sciacca, E. and
                 Vitello, F. and 
                 Colonnelli, I. and 
                 Becerra, Y. and
                 Cintero, E. S.  and
                 Marin, G.  and 
                 Jaros, M.  and 
                 Riha, L. and 
                 Strakos, P. and
                 Trujillo-Gomez, S. and 
                 Tramontana, E. and 
                 Wissing, R.},
        title = "{High Performance Visualization for Astrophysics and Cosmology}",
      journal = {PDP 2025 Conference Proceedings},
         year = {to appear}
}

@article{sciacca2015integrated,
  title={An integrated visualization environment for the virtual observatory: Current status and future directions},
  author={Sciacca, E. and Becciani, U. and Costa, A. and Vitello, F. and Massimino, P. and Bandieramonte, M. and Krokos, M. and Riggi, S. and Pistagna, C. and Taffoni, G.},
  journal={Astronomy and Computing},
  volume={11},
  pages={146--154},
  year={2015},
  publisher={Elsevier}
}

@misc{mignone2024,
      title={A Fourth-Order Finite Volume Scheme for Resistive Relativistic Magnetohydrodynamics}, 
      author={A. Mignone and V. Berta and M. Rossazza and M. Bugli and G. Mattia and L. Del Zanna and L. Pareschi},
      year={2024},
      eprint={2407.08519},
      archivePrefix={arXiv},
      primaryClass={astro-ph.HE},
      url={https://arxiv.org/abs/2407.08519}, 
}

@article{Berta2024b,
    author = {V. Berta and A. Mignone and M. Bugli and G. Mattia and M. Rossazza},
    title     = {Towards 4th-order accurate 3D Magnetic Reconnection in Relativistic Plasmas},
    journal   = {submitted to Astronum Proceedings},
    year      = {2024}
}

@article{berta_2024a,
title = {A 4th-order accurate finite volume method for ideal classical and special relativistic MHD based on pointwise reconstructions},
journal = {Journal of Computational Physics},
volume = {499},
pages = {112701},
year = {2024},
issn = {0021-9991},
doi = {https://doi.org/10.1016/j.jcp.2023.112701},
url = {https://www.sciencedirect.com/science/article/pii/S0021999123007969},
author = {V. Berta and A. Mignone and M. Bugli and G. Mattia}
}

@ARTICLE{spherinator,
       author = {{Polsterer}, K. L. and {Doser}, B. and {Fehlner}, A. and {Trujillo-Gomez}, S.},
        title = "{Spherinator and HiPSter: Representation Learning for Unbiased Knowledge Discovery from Simulations}",
      journal = {arXiv e-prints},
         year = 2024,
        month = jun,
          eid = {arXiv:2406.03810},
        pages = {arXiv:2406.03810},
          doi = {10.48550/arXiv.2406.03810},
archivePrefix = {arXiv},
       eprint = {2406.03810},
 primaryClass = {astro-ph.IM},
       adsurl = {https://ui.adsabs.harvard.edu/abs/2024arXiv240603810P}
}

@article{Porth_2017,
	doi = {10.1186/s40668-017-0020-2},
	year = 2017,  
	publisher = {Springer Science and Business Media {LLC}},
    url = "https://doi.org/10.1186%2Fs40668-017-0020-2",
	volume = {4},  
	author = {O. Porth and H. Olivares and Y. Mizuno and Z. Younsi and L. Rezzolla and M. Moscibrodzka and H. Falcke and M. Kramer},
	title = {The {B}lack {H}ole {A}ccretion {C}ode},
	journal = {Computational Astrophysics and Cosmology}
}

@article{Olivares_2019,
	doi = {10.1051/0004-6361/201935559},
	year = 2019,
	publisher = {{EDP} Sciences},
    url = "https://doi.org/10.1051%2F0004-6361%2F201935559",
	volume = {629},
	pages = {A61},
	author = {H. Olivares and O. Porth and J. Davelaar and E. R. Most and C. M. Fromm and Y. Mizuno and Z. Younsi and L. Rezzolla},
	title = {Constrained transport and adaptive mesh refinement in the Black Hole Accretion Code},
	journal = {Astronomy {\&} Astrophysics}
}

@article{Younsi_2020,
   title={Modelling the polarised emission from black holes on event horizon-scales},
   volume={14},
   ISSN={1743-9221},
   url={http://dx.doi.org/10.1017/S1743921318007263},
   DOI={10.1017/s1743921318007263},
   number={S342},
   journal={Proceedings of the International Astronomical Union},
   publisher={Cambridge University Press (CUP)},
   author={Younsi, Z. and Porth, O. and Mizuno, Y. and Fromm, C. M. and Olivares, H.},
   year={2020},
   month=apr, pages={9–12} }

@article{cfd_vis_workflow,
	title = {Workflow for high-quality visualisation of large-scale {CFD} simulations by volume rendering},
	volume = {200},
	issn = {09659978},
	url = {https://linkinghub.elsevier.com/retrieve/pii/S0965997824002291},
	doi = {10.1016/j.advengsoft.2024.103822},
	language = {en},
	urldate = {2024-11-29},
	journal = {Advances in Engineering Software},
	author = {Faltýnková, M. and Meca, O. and Brzobohatý, T. and Říha, L. and Jaroš, M. and Strakoš, P.},
	month = feb,
	year = {2025},
	pages = {103822},
}

@article{StreamFlow,
    author  = {I. Colonnelli and B. Cantalupo and I. Merelli and M. Aldinucci},
    doi     = {10.1109/TETC.2020.3019202},
    journal = {{IEEE} {T}ransactions on {E}merging {T}opics in {C}omputing},
    title   = {{StreamFlow}: cross-breeding cloud with {HPC}},
    url     = {https://doi.org/10.1109/TETC.2020.3019202},
    volume  = {9},
    number  = {4},
    pages   = {1723-1737},
    year    = {2021}
}

@ARTICLE{Teyssier2002,
       author = {{Teyssier}, R.},
        title = "{Cosmological hydrodynamics with adaptive mesh refinement. A new high resolution code called RAMSES}",
      journal = {Astronomy \& Astrophysics},
         year = 2002,
        month = apr,
       volume = {385},
        pages = {337-364},
          doi = {10.1051/0004-6361:20011817},
archivePrefix = {arXiv},
       eprint = {astro-ph/0111367},
 primaryClass = {astro-ph},
       adsurl = {https://ui.adsabs.harvard.edu/abs/2002A&A...385..337T}
}

@article{COLONNELLI2024,
title = {Cross-Facility Federated Learning},
journal = {Procedia Computer Science},
volume = {240},
pages = {3-12},
year = {2024},
note = {Proceedings of the First EuroHPC user day},
issn = {1877-0509},
doi = {https://doi.org/10.1016/j.procs.2024.07.003},
url = {https://www.sciencedirect.com/science/article/pii/S1877050924016909},
author = {I. Colonnelli and R. Birke and G. Malenza and G. Mittone and A. Mulone and J. Galjaard and L. Y. Chen and S. Bassini and G. Scipione and J. Martinovič and V. Vondrák and M. Aldinucci}
}

@inproceedings{ChaNGa1,
  author = "P. Jetley and F. Gioachin and C. Mendes and L. V. Kale and T. R. Quinn",
  title = "{Massively parallel cosmological simulations with ChaNGa}",
  booktitle = "Proceedings of IEEE International Parallel and Distributed Processing Symposium 2008",
  year = "2008",
}

@ARTICLE{ChaNGa2,
       author = {{Menon}, H. and {Wesolowski}, L. and {Zheng}, G. and {Jetley}, P. and {Kale}, L. and {Quinn}, T. and {Governato}, F.},
        title = "{Adaptive techniques for clustered N-body cosmological simulations}",
      journal = {Computational Astrophysics and Cosmology},
         year = 2015,
        month = mar,
       volume = {2},
          eid = {1},
        pages = {1},
          doi = {10.1186/s40668-015-0007-9},
archivePrefix = {arXiv},
       eprint = {1409.1929},
 primaryClass = {astro-ph.IM},
       adsurl = {https://ui.adsabs.harvard.edu/abs/2015ComAC...2....1M}
}

@ARTICLE{ChaNGaSPH,
       author = {{Wadsley}, J. W. and {Keller}, B. W. and {Quinn}, T. R.},
        title = "{Gasoline2: a modern smoothed particle hydrodynamics code}",
      journal = {Monthly Notices of the Royal Astronomical Society},
         year = 2017,
        month = oct,
       volume = {471},
       number = {2},
        pages = {2357-2369},
          doi = {10.1093/mnras/stx1643},
archivePrefix = {arXiv},
       eprint = {1707.03824},
 primaryClass = {astro-ph.IM},
       adsurl = {https://ui.adsabs.harvard.edu/abs/2017MNRAS.471.2357W}
}

@PHDTHESIS{ChaNGaGrav,
       author = {{Stadel}, J. G.},
        title = "{Cosmological N-body simulations and their analysis}",
       school = {University of Washington, Seattle},
         year = 2001,
        month = jan,
       adsurl = {https://ui.adsabs.harvard.edu/abs/2001PhDT........21S}
}

@ARTICLE{Groth23,
       author = {{Groth}, F. and {Steinwandel}, U. P. and {Valentini}, M. and {Dolag}, K.},
        title = "{The cosmological simulation code OPENGADGET3 - implementation of meshless finite mass}",
      journal = {Monthly Notices of the Royal Astronomical Society},
         year = 2023,
        month = nov,
       volume = {526},
       number = {1},
        pages = {616-644},
          doi = {10.1093/mnras/stad2717},
archivePrefix = {arXiv},
       eprint = {2301.03612},
 primaryClass = {astro-ph.IM},
       adsurl = {https://ui.adsabs.harvard.edu/abs/2023MNRAS.526..616G}
}

@ARTICLE{Springel2005,
       author = {{Springel}, V.},
        title = "{The cosmological simulation code GADGET-2}",
      journal = {Monthly Notices of the Royal Astronomical Society},
         year = 2005,
        month = dec,
       volume = {364},
       number = {4},
        pages = {1105-1134},
          doi = {10.1111/j.1365-2966.2005.09655.x},
archivePrefix = {arXiv},
       eprint = {astro-ph/0505010},
 primaryClass = {astro-ph},
       adsurl = {https://ui.adsabs.harvard.edu/abs/2005MNRAS.364.1105S}
}

@ARTICLE{Beck2016,
       author = {{Beck}, A.~M. and {Murante}, G. and {Arth}, A. and {Remus}, R. -S. and {Teklu}, A.~F. and {Donnert}, J.~M.~F. and {Planelles}, S. and {Beck}, M.~C. and {F{\"o}rster}, P. and {Imgrund}, M. and {Dolag}, K. and {Borgani}, S.},
        title = "{An improved SPH scheme for cosmological simulations}",
      journal = {Monthly Notices of the Royal Astronomical Society},
         year = 2016,
        month = jan,
       volume = {455},
       number = {2},
        pages = {2110-2130},
          doi = {10.1093/mnras/stv2443},
archivePrefix = {arXiv},
       eprint = {1502.07358},
 primaryClass = {astro-ph.CO},
       adsurl = {https://ui.adsabs.harvard.edu/abs/2016MNRAS.455.2110B}
}

@ARTICLE{Ragagnin2020,
       author = {{Ragagnin}, A. and {Dolag}, K. and {Wagner}, M. and {Gheller}, C. and {Roffler}, C. and {Goz}, D. and {Hubber}, D. and {Arth}, A.},
        title = "{Gadget3 on GPUs with OpenACC}",
      journal = {arXiv e-prints},
         year = 2020,
        month = mar,
          eid = {arXiv:2003.10850},
        pages = {arXiv:2003.10850},
          doi = {10.48550/arXiv.2003.10850},
archivePrefix = {arXiv},
       eprint = {2003.10850},
 primaryClass = {astro-ph.IM},
       adsurl = {https://ui.adsabs.harvard.edu/abs/2020arXiv200310850R}
}
\end{document}